\documentclass[aps,prl,citeautoscript,reprint,superscriptaddress,floatfix,footinbib,showkeys]{revtex4-1}
\usepackage{setspace}
\usepackage[utf8]{inputenc}
\usepackage[colorlinks=true,linkcolor=black,urlcolor=black,filecolor=black,citecolor=black]{hyperref}
\usepackage{color,soul}
\usepackage{amsmath}
\usepackage{amsfonts}
\usepackage{amssymb}
\usepackage{graphicx}
\usepackage{dcolumn}
\usepackage{bm}
\usepackage{subfigure}
\usepackage{booktabs}
\usepackage{siunitx}
\usepackage[normalem]{ulem}

\newcommand{\Gbulk}{G_\mathrm{bulk}}
\newcommand{\kB}{k_\mathrm{B}}

\newcommand{\FDP}{\textcolor{black}}
\definecolor{darkgreen}{rgb}{0,0.6,0}
\newcommand{\OM}{\textcolor{black}}
\newcommand{\AZ}{\textcolor{black}}
\begin{document}

\title{Non-universal localization transition in the quantum Hall effect probed through broken-symmetry states of graphene %
}

\author{Aifei Zhang}
\affiliation{Université Paris-Saclay, CEA, CNRS, SPEC, 91191 Gif-sur-Yvette, France} 
\author{Torsten R\"oper}
\affiliation{II. Physikalisches Institut, Universit\"at zu K\"oln, Z\"ulpicher Str. 77, D-50937 K\"oln, Germany}
\author{Manjari Garg}
\affiliation{Université Paris-Saclay, CEA, CNRS, SPEC, 91191 Gif-sur-Yvette, France} 
\affiliation{Department of Physics, Indian Institute of Technology Roorkee, Uttarakhand 247667, India} 
\author{Kenji Watanabe}
\affiliation{Research Center for Electronic and Optical Materials, National Institute for Materials Science, 1-1 Namiki, Tsukuba 305-0044, Japan} 
\author{Takashi Taniguchi} 
\affiliation{Research Center for Materials Nanoarchitectonics, National Institute for Materials Science,  1-1 Namiki, Tsukuba 305-0044, Japan}
\author{Carles Altimiras} \affiliation{Université Paris-Saclay, CEA, CNRS, SPEC, 91191 Gif-sur-Yvette, France} 
\author{Patrice Roche} 
\affiliation{Université Paris-Saclay, CEA, CNRS, SPEC, 91191 Gif-sur-Yvette, France} 
\author{Erwann Bocquillon}
\affiliation{II. Physikalisches Institut, Universit\"at zu K\"oln, Z\"ulpicher Str. 77, D-50937 K\"oln, Germany}
\author{Olivier Maillet}\email{olivier.maillet@cea.fr}
\affiliation{Université Paris-Saclay, CEA, CNRS, SPEC, 91191 Gif-sur-Yvette, France}
\author{François D. Parmentier}\email{francois.parmentier@phys.ens.fr} \affiliation{Université Paris-Saclay, CEA, CNRS, SPEC, 91191 Gif-sur-Yvette, France} \affiliation{Laboratoire de Physique de l’Ecole normale sup\'erieure, ENS, Universit\'e PSL, CNRS, Sorbonne Universit\'e, Universit\'e Paris Cit\'e, F-75005 Paris, France } 

\date{\today}

\begin{abstract}

The quantum Hall effect hosts quantum phase transitions in which the localization length, that is the size of disorder-induced bulk localized states, is governed by universal scaling from percolation theory. However, this universal character is not systematically observed in experiments, including very recent ones in extremely clean devices. Here we explore this non-universality by systematically measuring the localization length in broken-symmetry quantum Hall states of graphene. Depending on the nature and gap size of these states, we observe differences of up to a tenfold in the minimum localization length, accompanied by clear deviations from universal scaling. Our results, as well as the previously observed non-universality, are fully captured by a simple picture based on the co-existence of localized states from two successive sub-Landau levels.

\end{abstract}

\maketitle

The Quantum Hall effect (QHE) is characterized by the emergence at low temperature of an insulating bulk, and of quantized edge channels which ballistically transport current along the edges of the system. While this essentially stems from the quantization of an otherwise continuous electronic spectrum into Landau levels (LLs), disorder turns out to play an essential role. Indeed, without disorder, the QHE only develops at precise ratios of the carrier density $n_e$ and the magnetic field $B$, corresponding to integer values of the filling factor $\nu=n_e h/eB$, where $h$ is Planck's constant and $e$ the electron charge. Remarkably, adding disorder stabilizes the QHE, leading to plateaus in both $B$ and $n_e$ where quantization remains enforced even though $\nu$ is not integer. This is due to the presence of bulk localized states that do not participate in quantized transport along the edge, while being able to accommodate the superfluous charge carriers. These localized states correspond to the disorder-broadened part of a LL, while its center host extended states for which bulk transport is restored~\cite{Goerbig2009}. Moving the Fermi level from one type of state to another gives rise to a \textit{localization transition} which has been extensively investigated since the very first experimental evidences of the QHE. On the theory side, this transition is characterized by universal scalings~\cite{Huckestein1995}; in particular, the localization length (hereafter denoted $\xi$), corresponding to the average size of the loops formed by extended/localized states, is expected to be a power law of the filling factor~\cite{Pruisken1988,Huckestein1990, Huckestein1995,Fogler1997}:

\begin{equation} \label{eq_scaling}
    \xi \propto|\nu-\nu_c|^{-\gamma},
\end{equation}

 with an universal exponent $\gamma\approx2.35$~\cite{Huckestein1990,Rimon2026}, and where $\nu_c$ is the critical value of the filling factor where electrons are fully delocalized. Thus, $\xi$ is expected to change strongly with $\nu$, with variations over several orders of magnitude between minima comparable to the magnetic length (of a few nm), and maxima comparable to the sample size, typically in the $10~$\unit{\um} range. However, experiments have shown significant discrepancies~\cite{Wei1988,Furlan1998,Giesbers2009, Bennaceur2012,Dodoo-Amoo2014,Arapov2019,Madathil2023,Kaur2023,Yeh2024}, with the universal scaling exponent $\gamma\approx2.35$ being generally observed only in very clean samples (including in the fractional QHE~\cite{Engel1990,Madathil2023,Kaur2023}), and reported minima of $\xi$ being sometimes up to 100 times larger than expected~\cite{Bennaceur2012}. Systematic experiments in GaAs/GaAlAs heterostructures~\cite{Wei1992,Li2005,Li2009} suggested that the observed scaling exponent depends on the type of disorder, with short-range disorder yielding the predicted values, and measurements in spin/valley degenerate graphene LLs reported observations corresponding to $\gamma$ markedly higher than $2.35$~\cite{Hwang1993,Peters2014}.

\begin{figure*}[hbt!]
    \centering
    \includegraphics[width=0.95\linewidth]{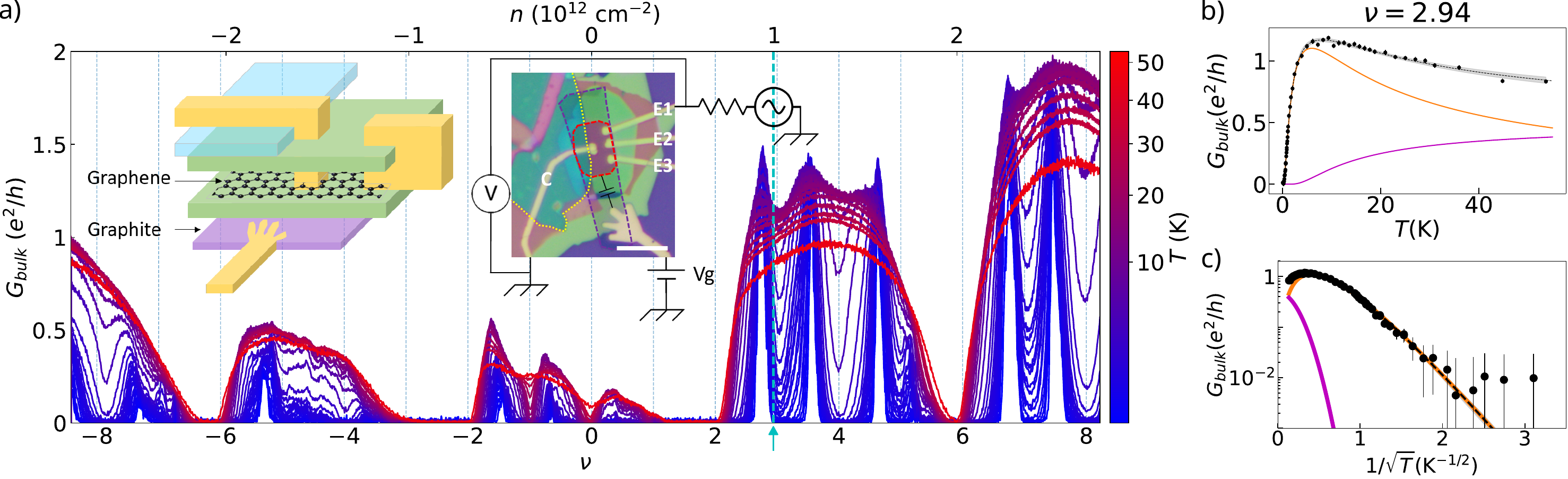}
    \caption{(a) Bulk conductance $\Gbulk$ measured at $B=14\,$T as a function of filling factor, for temperatures ranging from $50 \, \mathrm{mK}$ (blue) to $\approx 55\, \mathrm{K}$ (red). Insets: left: Sample schematic. hBN layers encapsulating the MLG flake are shown in green,  the graphite gate in purple, and the topmost hBN layer separating the central contact from the edges in blue. Right: sample optical micrograph and measurement circuit. Red dashed line: MLG flake; purple dashed line: graphite gate; yellow dotted line: topmost hBN. Scale bar: $10~$\unit{\um}. \FDP{(b) and (c) Linecut at $\nu=2.94$ (cyan vertical dashed line in (a)) versus $T$ (b) and $1/\sqrt{T}$ (c). Symbols: experimental data. Dashed line: fit combining VRH (orange - $T_0=23.8~$K, $G_0=48.4~\left[e^2/h\right]/K$) and Arrhenius activation (magenta - $\Delta^\ast/\kB=54.0~$K, $G_1=0.49~e^2/h$).}}
    
    \label{fig:fig1}
\end{figure*}

We have addressed this issue by probing the temperature dependence of QH states in graphene in a Corbino geometry \cite{Peters2014,Zeng2019}, over three orders of magnitude. The fourfold spin and valley symmetry of graphene is lifted at high magnetic field \cite{Nomura2006}, providing well-separated scales for the energy level spacing \cite{Young2012}: cyclotron gap for a fully filled LL, and spin/valley gaps for the symmetry broken sub-LLs. Tuning the carrier density with an electrostatic gate allows us to probe states with vastly different gaps at otherwise fixed magnetic length and disorder. We show that there is a clear link between the type of the sub-LL gaps and the measured localization length $\xi$, as well as the exponent $\gamma$. We clarify this link through a picture based on the overlap of localized states from different successive sub-LLs, preserving the plateau quantization at $T=0~$K but rendering it more fragile at finite temperature. This picture allows explaining most of the discrepancies observed in past experiments.

Our devices consist of monolayer graphene (MLG) flakes encapsulated by top and bottom hexagonal boron nitride (hBN) flakes (see insets of Fig. \ref{fig:fig1}). We change the filling factor by tuning the carrier type and density with a graphite gate, separated from the MLG by the $\approx50~$nm-thick bottom hBN flake. To probe exclusively the transport in the bulk of graphene, we employ a Corbino geometry \cite{Zeng2019}, with one 1D side contact~\cite{Wang2013} located in the center of the device (C), and three other (E1, E2, E3) along the edge. The lead connecting the central contact is isolated from the graphene edge by another flake of hBN deposited onto the device after the central contact was made. We measure the bulk conductance $\Gbulk$ at high magnetic field in a dilution refrigerator using a low frequency ($<10$ Hz) lock-in technique, while varying the filling factor and the temperature $T$. The main panel of Fig.~\ref{fig:fig1} shows the dependence of $\Gbulk$ with $\nu$ measured at $B=14~$T, for $T$ ranging from $50~$mK (blue) to $55~$K (red). At low temperature, $\Gbulk$ shows clear zeroes at all integer $\nu$, signaling well-developed QH states for cyclotron gaps ($\nu=\pm2,\pm6$) and symmetry-broken states, with sharp peaks at transitions between QH states. Note that at low $T$, some  peaks do not appear (especially for negative filling factors), which we attribute to the formation of a p-n junction around the central contact~\cite{SM,Zhao2025}. Increasing the temperature leads to finite $\Gbulk$ on the plateaus, with a non-monotonous behavior as a function of $T$ particularly visible in the $N=1$ and $N=2$ LLs. Plateaus originating from spin and valley symmetry breakings are destroyed first while cyclotron gaps are much more robust, with $\nu\pm2$ remaining largely quantized at $55~$K.

\FDP{The non-monotonous behavior of $\Gbulk$ with  $T$ is illustrated in Fig.~\ref{fig:fig1}(b) for $\nu=2.94$, with a sharp increase at very low temperature, and a slow decrease above $15~$K. This variation can be fitted with a function combining Efros-Shklovskii variable range hopping (ES-VRH)~\cite{Shklovskii1984,Shklovskii2024} (\AZ{see also Supplemental Material Section IV}\cite{SM}), which describes bulk conduction through localized states at low temperature, and Arrhenius thermal activation at higher temperature:}

\begin{equation} \label{eq_Gfit}
    \Gbulk(T) = \frac{G_0}{T} \exp{\left[-\sqrt{\frac{T_0}{T}}\right]}+G_1\exp{\left[-\frac{\Delta^\ast}{2\kB T}\right]}.
\end{equation}
\FDP{The phenomenological prefactor $G_0/T$ in the VRH contribution is associated to electron-phonon coupling~\cite{Furlan1998}}, $\Delta^\ast$ is the thermal activation energy, and $\kB T_0$ is the characteristic hopping energy of the ES-VRH model. $T_0$ is linked to the localization length $\xi$ (physically representing the average size of localized states) through the following relation of the ES-VRH model~\cite{Efros1979,Shklovskii2024}: 

\begin{equation} \label{eq_loc_length}
    \xi = \frac{Ce^2}{4\pi\epsilon k_B T_0},
\end{equation} 
where $C\approx 6.2$~\cite{Shklovskii2024}, $e$ is the electron charge, and $\epsilon$ is the dielectric constant of hBN. \FDP{ Importantly, the different energy scales at play in VRH and thermal activation allow a clear separation of the two contributions, with VRH being largely dominant at low temperature, as illustrated by Fig.~\ref{fig:fig1}(c). This behavior (which we systematically fit for all $\nu$, see below) allows us to test the universality of the low-temperature dependence of $\Gbulk$ by using the scaling introduced in ref.~\cite{Hohls2002}. Considering the ES-VRH term of Eq.~\ref{eq_Gfit} along with the universal power law of $\xi/T_0$ of Eqs.~\ref{eq_scaling} and~\ref{eq_loc_length} suggests the rescaled variable $s=|\nu-\nu_c|^{\gamma}/T\propto T_0/T$ universally seting the dependence of $\Gbulk(\nu,T)$ near a given $\nu_c$. The universal scaling then appears as the collapse of all data upon a single exponential law when plotting $\Gbulk/s$ versus $s^{1/2}$~\cite{Hohls2002}. Systematically testing this scaling on our data for $T<5~\mathrm{K}$ yields a remarkable observation, illustrated in Fig.~\ref{fig:Hohlsscaling} for the $\nu=2\leftrightarrow3$ and $\nu=6\leftrightarrow7$ transitions (see Supplemental Material Section II.B~\cite{SM} for the other transitions). The critical filling factor $\nu_\mathrm{c}$ is defined at the low-temperature maxima of $\Gbulk$, see End Matter. While the data corresponding to cyclotron gaps (left panel in Fig.~\ref{fig:Hohlsscaling}) agree well with a collapse over more than 2 orders of magnitude in $\Gbulk/s$ with the predicted universal scaling exponent $\gamma=2.35$, the scaling fails for data from symmetry-broken gaps (right panel in Fig.~\ref{fig:Hohlsscaling}). This discrepancy suggests that, at first order, the magnitude of the gap plays a role in the observation of the predicted universal scaling.}

\begin{figure}[hbt!]
    \centering
    \includegraphics[width=0.95\linewidth]{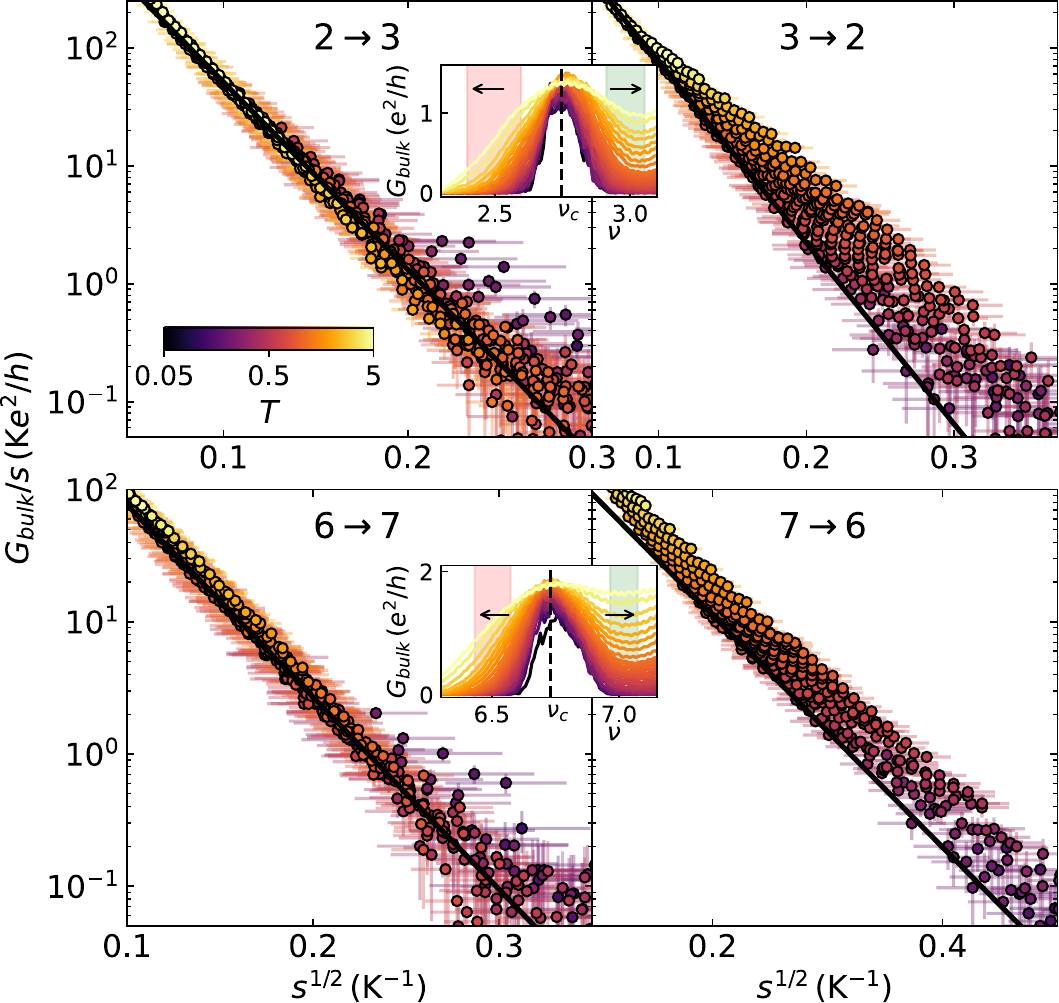}
   \caption{\FDP{Rescaled conductance $\Gbulk/s$ (symbols) versus $s^{1/2}$ at the $\nu=2\leftrightarrow3$ (top) \OM{and $6\leftrightarrow7$ (bottom)} transitions for $T<5~$K. Inset: raw data, indicating the position of $\nu_c=2.745$ for \OM{the $2\leftrightarrow 3$} transition (resp. $\nu_c=6.731$ for \OM{the $6\leftrightarrow 7$} transition). The shaded areas indicate the range of data shown in the main panel. Left: $\nu<\nu_c$, corresponding to the end of the $\nu=2$ (resp. $\nu=6$) plateau. Right: $\nu>\nu_c$, corresponding to the start of the $\nu=3$ (resp. $\nu=7$) plateau. The symbols colors correspond to the temperature, indicated by the color bar on the left panel. Full lines: predicted universal scaling with $\gamma=2.35$. \OM{X-axis error bars correspond to the uncertainty in the definition of $\nu_c$. Y-axis error bars combine this uncertainty and the measurement noise.}}}
    \label{fig:Hohlsscaling}
\end{figure}

\begin{figure}[hbt!]
    \centering
    \includegraphics[width=0.9\linewidth]{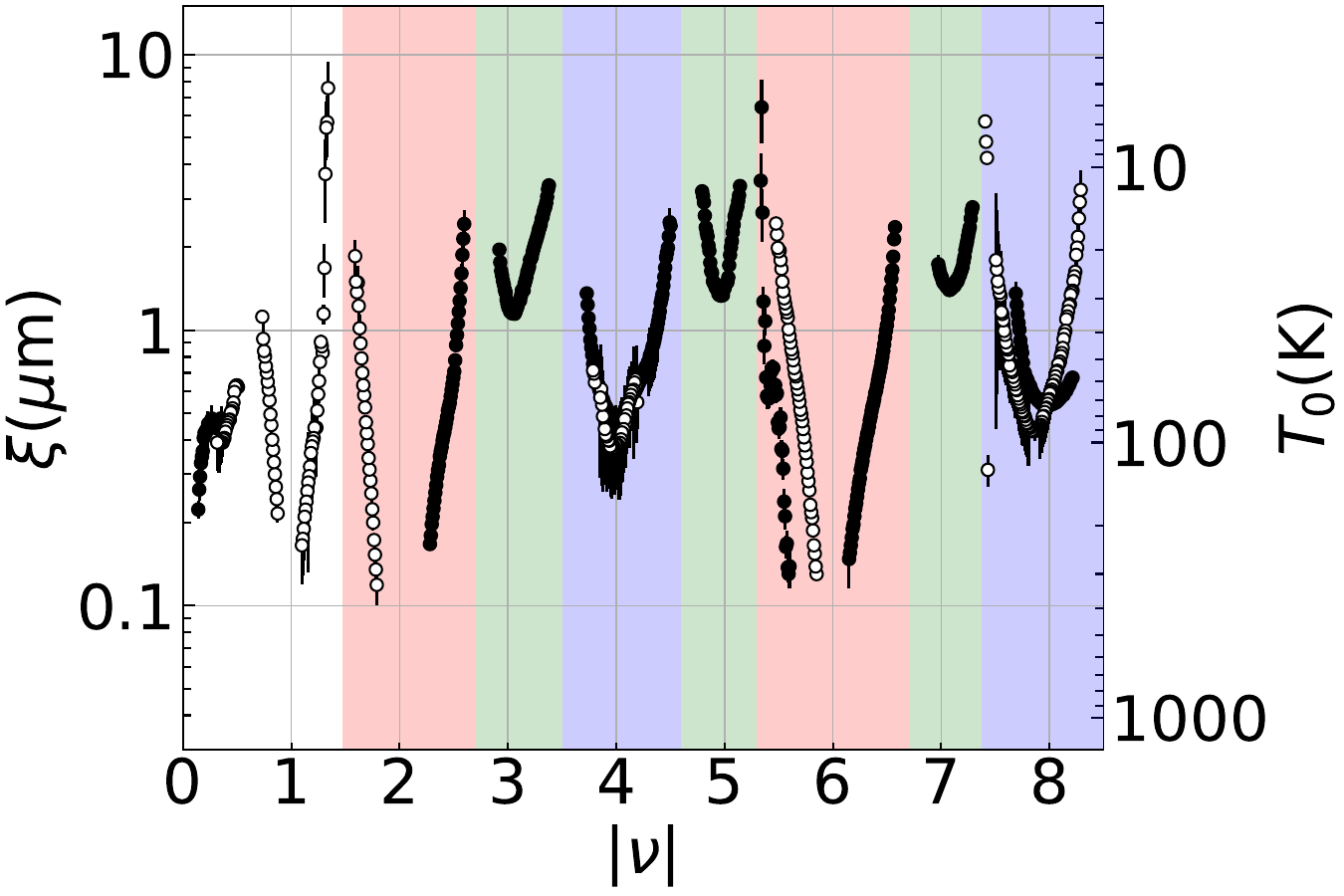}
   \caption{Localization length $\xi$ (left $Y$-axis) and corresponding hopping temperature $T_0$ (right $Y$-axis) extracted from the fits, plotted in log scale versus $\vert\nu\vert$. Full symbols: positive $\nu$ (electron doping). Open symbols: negative $\nu$ (hole doping). The background color indicates the nature of the gaps for $\nu\geq2$. }
    \label{fig:loc_length}
\end{figure}

\FDP{To test this hypothesis, we repeat the fitting procedure using Eq.~\ref{eq_Gfit} shown in Fig.~\ref{fig:fig1}(b) and (c) for all of our data. This allows extracting $\xi$ for every swept values of $\nu$ shown in Fig.~\ref{fig:fig1}(a). Note that we have checked that the extracted $\xi$ do not change upon fitting the data below $2~$K with only the ES-VRH part of Eq.~\ref{eq_Gfit} (see Supplemental Material Section II.D~\cite{SM}); we have also checked that our data matches better with ES-VRH than with Mott-VRH~\cite{SM}. The result is shown in Fig.~\ref{fig:loc_length}, plotting the obtained $\xi$ and their corresponding $T_0$ as a function of $\nu$, for both electron (full symbols) and hole (open symbols) doping. Marked quantitative differences in $\xi(\nu)$ are observed depending on the type of QH gap.} Cyclotron gaps ($\nu=\pm2,\pm6$, red shading) show a very strong reduction of $\xi$, from a few $\mu$m, comparable to sample size, down to approximately $100~$nm close to the center of the plateau. \FDP{Smaller values of $\xi$ (down to $\xi_\mathrm{min}\approx20~$nm) near the center of cyclotron gaps can be extracted when restricting the fits to ES-VRH only (see Supplemental Material Section~II.E).} Spin-polarized~\cite{Young2012} ($\nu=\pm4,\pm8$, blue shading) and valley-polarized~\cite{Young2012} ($\nu=3,5,7$, green shading) gaps show much larger minima: $\xi_\mathrm{min}\approx500~$nm for spin gaps, and $\xi_\mathrm{min}\approx1~$\unit{\um} for valley gaps. We stress the qualitative and quantitative similarities between gaps of the same nature, for both types of carrier doping: as the carrier density is changed along a given QH plateau, $\xi$ decreases from a value comparable to the size of the sample at the edges of the plateau where electrons are delocalized, and reaches a minimal value at the center of the plateau that reflects the degree of localization of each QH state. Note that the data corresponding to partial filling of the $N=0$ LL ($\nu=-1,0$, no shading) stands out from the rest, with low $\xi_\mathrm{min}\approx100~$nm, despite the fact that these are symmetry-broken QH states. The $N=0$ LL is expected to have a smaller disorder broadening than higher LLs~\cite{Giesbers2007,Malla2019}, in part due to the fact that the wave function at $N=0$ does not mix orbitals from other LLs. Because of this singular behavior, we analyze the corresponding data separately in the Supplemental Material~\cite{SM}, and focus on higher LLs in the rest of this Letter.

\begin{figure}[hbt!]
    \centering
    \includegraphics[width=0.9\linewidth]{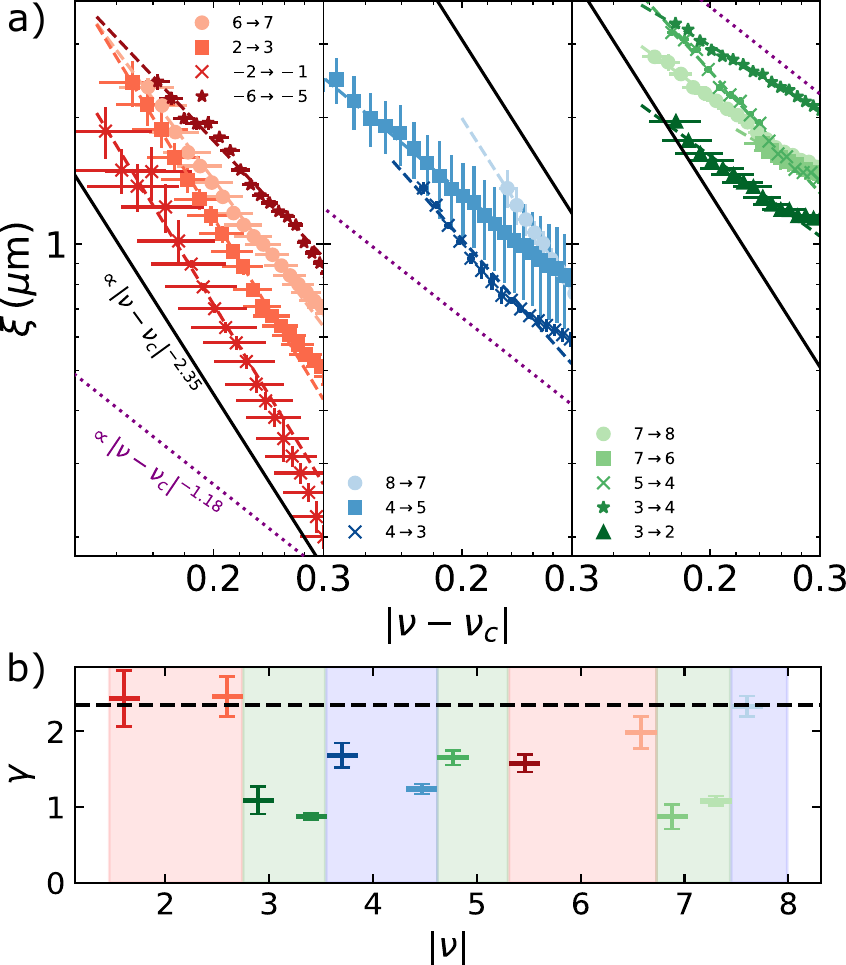}
    \caption{ \FDP{(a),} Localization length $\xi$  versus $|\nu-\nu_c|$ (see text), for different transitions between quantum Hall gaps. Left panel: transitions from cyclotron gaps towards symmetry broken gaps. Center panel: transitions from spin gaps towards valley gaps. Right panel: transitions from valley gaps towards cyclotron/spin gaps. \FDP{Dashed lines: power law fits. Black line: power law with the universal scaling exponent of $\gamma=-2.35$. Purple dotted line: power law with $-1.18$ exponent. (b), exponents $\gamma$ obtained from the fits in (a) versus $\vert \nu \vert$. The lines and background color indicate the nature of the gaps.}}
    \label{fig:scaling}
\end{figure}

We compare our extracted $\xi$ with the predicted power law of Eq.~\ref{eq_scaling} \FDP{by plotting it versus $|\nu-\nu_c|$, for $|\nu-\nu_c|<0.3$.} Fig.~\ref{fig:scaling}(a) shows the corresponding plots for the three different types of gaps: cyclotron (left panel, red-hue symbols), spin (middle panel, blue-hue symbols), and valley (right panel, green-hue symbols). The solid black line indicates the universal scaling exponent $\gamma = 2.35$. \FDP{Fitting the data with a power law (dashed lines) yields the exponents shown in Fig.~\ref{fig:scaling}(b), plotted versus $\vert\nu\vert$ and highlighted according to the nature of the gap. Cyclotron gaps (red) tend to match the universal scaling, while symmetry-broken gaps tend to have smaller exponents, closer to half the universal scaling exponent (taken as a guide to the eye, see purple dotted lines in Fig.~\ref{fig:scaling}(a)). Note that using the extracted values of $\gamma$ shown in Fig.~\ref{fig:scaling}(b) to plot the rescaled conductances $\Gbulk/s$ vs $s^{1/2}$ leads to a convincing collapse of the data below $5~$K, as shown in Supplemental Material Section II.B~\cite{SM}.}

\begin{figure}[hbt!]
    \centering
    \includegraphics[width=0.9\linewidth]{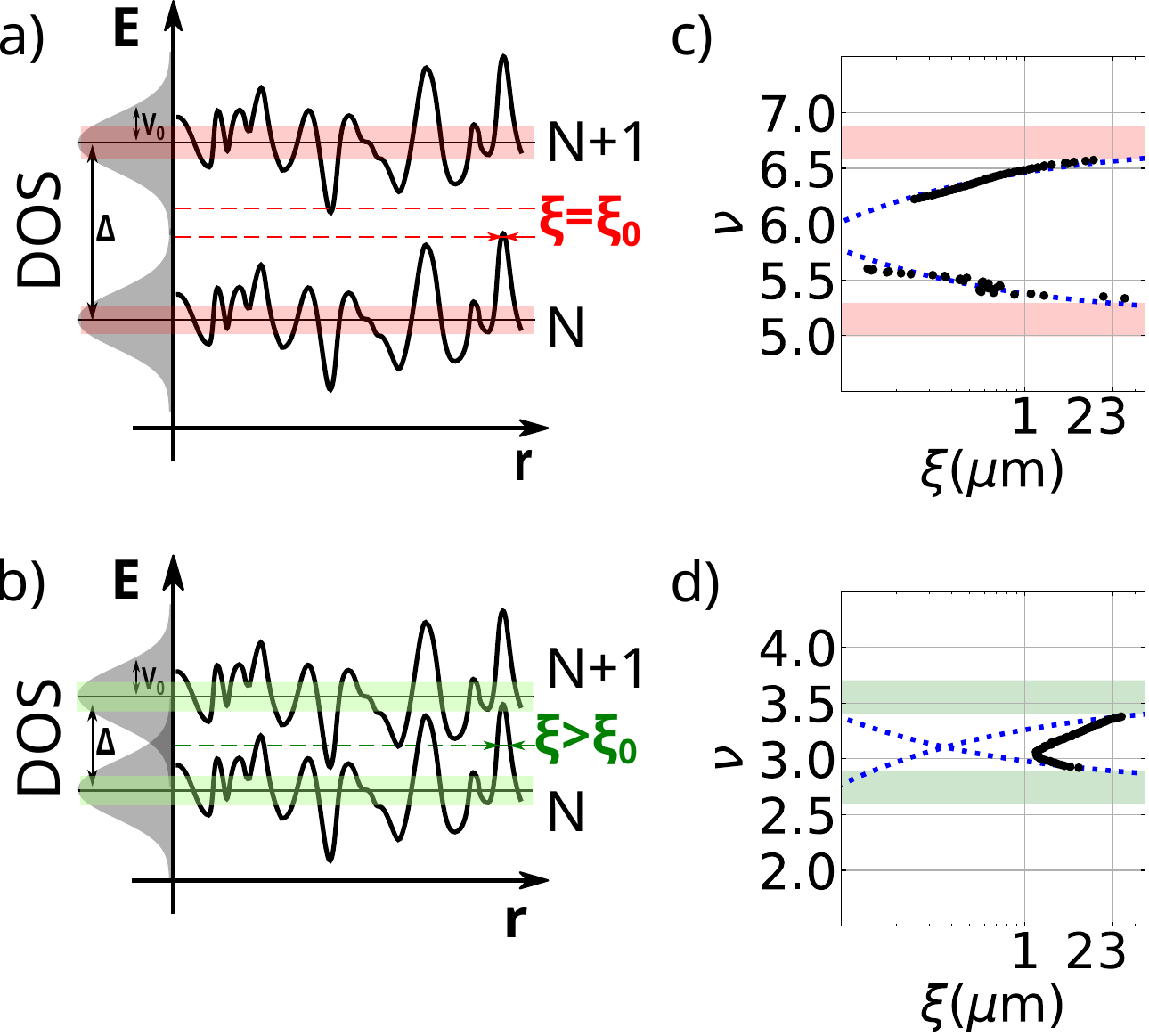}
    \caption{
    \FDP{(a) and (b),} Sketches of the typical sub-LLs spacing and disorder along the spatial coordinate $r$ for cyclotron (a) and valley (b) gaps. Horizontal dashed lines: position of the Fermi level at the center of the QH plateau, with the corresponding typical $\xi$ indicated by the arrows. Thick shaded horizontal lines: extended states. \FDP{(c) and (d),} Typical $\xi$ versus $\nu$ plots taken from Fig.~\ref{fig:loc_length}, for the $\nu=6$ cyclotron gap (c) and the $\nu=3$ valley gap (d). Shaded lines: typical position of the extended states. \AZ{Blue dashed lines: Eq.~\ref{eq_scaling} with $\nu_c$ from Table~\ref{tab:critical_filling} and $\gamma=2.35$, indicating the $|\nu-\nu_c|^{-2.35}$ of Eq.~\ref{eq_scaling}.}}
        
    \label{fig:gap_vs_disorder}
\end{figure}

To understand the difference in the localization lengths minima and scaling exponents measured for the three types of gaps, we propose an intuitive picture comparing the size of the gaps $\Delta$ to the amplitude of the disorder $V_0$ that is randomly distributed in real space, and that in first approximation applies similarly to all sub-LLs. \FDP{This picture is represented in Fig.~\ref{fig:gap_vs_disorder}(a) and (b). The first case, depicted in Fig.~\ref{fig:gap_vs_disorder}(a), corresponds to cyclotrons gaps in our experiment, where $\Delta \gg V_0$. In this case, disorder-broadened sub-LLs are sufficiently distant in energy, and there is no overlap between the localized states of sub-LLs $N$ and $N+1$. As the density is swept along the corresponding QH plateau, the Fermi level (horizontal dashed lines in Fig.~\ref{fig:gap_vs_disorder}(a) and (b)) jumps from the top of sub-LL $N$ to the bottom of sub-LL $N+1$, and $\xi$ reaches its minimal theoretical value $\xi_0$ corresponding to the size of the last available localized state (arrows in Fig.~\ref{fig:gap_vs_disorder}(a)). Thus, the measured $\xi$ drops from values comparable to the sample size when the Fermi level is close to extended states (red and green horizontal bands in Fig.~\ref{fig:gap_vs_disorder}(c) and (d), respectively), to values smaller than $100~$nm in the center of the plateau. In the second case (Fig.~\ref{fig:gap_vs_disorder}(b)), corresponding to valley gaps in our experiments, $\Delta$ is comparable to $V_0$, such that there is significant overlap between the localized states of sub-LLs $N$ and $N+1$. In the middle of the plateau, the Fermi level thus intersects localized states of both sub-LLs, such that the decrease of $\xi$ accompanying the filling of sub-LL $N$ turns into an increase as the Fermi level reaches the bottom of sub-LL $N+1$ before sub-LL $N$ is fully filled. This yields significantly higher $\xi_\mathrm{min}\gg\xi_0$, as shown in the plot of Fig.~\ref{fig:gap_vs_disorder}(d) for the valley-polarized state at $\nu=3$.} Note that smaller ratios $\Delta/V_0$ lead to overlaps between localized and extended states of the two sub-LLs and a loss of quantization. Our picture of LL overlap also explains the observed scalings exponents. The universal scaling theory should only be observed when the localized states of different sub-LLs do not overlap; for overlapping sub-LLs, their opposite contributions decrease the dependence of $\xi(\nu)$, leading to smaller observed power law exponents. \FDP{This is illustrated in Figs.~\ref{fig:gap_vs_disorder}(c) and (d), where the blue dashed lines indicate the expected dependence of $\xi$ with the universal scaling, which strongly deviates from the data for the $\nu=3$ valley gap.} Our picture is consistent with the values of the activation energies extracted from the fits of Eq.~\ref{eq_Gfit} \FDP{(shown in Fig.~\ref{fig:activation_gap_14T_ES+activation} of the End Matter)} and the estimation of disorder broadening $V_0/\kB\approx40~$K obtained from low-field magnetotransport~\cite{SM}: for valley-polarized states, activation gap and disorder broadening are very comparable, while the former is expected to be markedly larger for spin and cyclotron gaps~\cite{Young2012}.

Note that in our proposed picture, the disorder is similar for all types of gaps in graphene. This is not generally the case, as different types of disorder (such as magnetic impurities, adatoms~\cite{Dutreix2019} or atomic defects~\cite{Joucken2021}) will not affect the spin and valley index of the electron wavefunctions in the same way. Nevertheless, as our argument is essentially based on the overlap between the localized states of disorder-broadened sub-LLs, it also holds for different types of disorder and different broadenings. Comparing the hopping energy normalized by the gap $\kB T_0/\Delta$ for valley and cyclotron gaps (taking the fitted value $\Delta^\ast$ for the former, and the standard prediction~\cite{Goerbig2011} for the latter), while reducing the discrepancies between the two gap types to within a factor 5, still shows clear differences (see Supplemental Material Section II.G~\cite{SM}), possibly suggesting a role of valley \FDP{(and, perhaps more generally, symmetry-broken)} specific disorder.

Finally, our picture allows explaining most of the discrepancies of previous experiments in which the disorder was not explicitly long-range~\cite{Li2005}. For instance, higher values of $\gamma$ observed in spin/valley degenerate LL~\cite{Hwang1993} can be understood by the added contributions of VRH transport through the localized states of two sub-LLs which are split in energy by an amount much smaller than their broadening. \FDP{There is thus an important need to clarify the nature of the extracted $\xi$, which only corresponds to the actual localization length in the limit of an isolated sub-LL, and should otherwise be interpreted as an effective hopping length.}

Note that we discuss the breakdown mechanisms at large zero-frequency and finite-frequency biases in a separate article~\cite{Roeper2025}.

\begin{acknowledgments}
We warmly thank F. Ladieu, B. Piot and M. O. Goerbig for stimulating discussions. This work was funded by the ERC (ERC-2018-STG \textit{QUAHQ}), by the “Investissements d’Avenir” LabEx PALM (ANR-10-LABX-0039-PALM), and by the Region Ile de France through the DIM QUANTIP. O.M. acknowledges funding from the ANR (ANR-24-CE47-2695 NONABEG, ANR-23-CE47-0002 CRAQUANT and ANR-25-ERCS-0006-01 ANYQUAC). K.W. and T.T. acknowledge support from the JSPS KAKENHI (Grant Numbers 21H05233 and 23H02052) , the CREST (JPMJCR24A5), JST and World Premier International Research Center Initiative (WPI), MEXT, Japan. 
\end{acknowledgments}

\section{End Matter}

\subsection{Extraction of the critical filling factor}
In the analysis shown here, $\nu_c$ is defined by the position of the low-temperature maxima of $\Gbulk$ between two plateaus in Fig.~\ref{fig:fig1}. The list of extracted $\nu_c$ is shown in Table~\ref{tab:critical_filling}; the error, corresponding to the uncertainty on the position of the minima, is propagated into the error bars of Figs.~\ref{fig:Hohlsscaling} and~\ref{fig:scaling}. The full extraction procedure is shown in Supplemental Section~II.A~\cite{SM}. Note that our results and their interpretation are robust to a change in the extraction procedure (\textit{e.g.} by taking the middle of a conductance peak)~\cite{SM}.

\begin{table}[h]
\centering
\begin{tabular}{|c|c|c|c|c|c|}
\hline
transition & $-8\leftrightarrow -7$ & $-6\leftrightarrow -5$ & $-2\leftrightarrow -1$ & $2\leftrightarrow 3$ \\

\hline
$\nu_c^\mathrm{max}$ & -7.338(22) & -5.278(8) & -1.455(25) & 2.745(17) \\

\hline
transition & $3\leftrightarrow 4$ & $4\leftrightarrow 5$ & $6\leftrightarrow 7$ &
 $7\leftrightarrow 8$ \\
 \hline
$\nu_c^\mathrm{max}$  & 3.555(3) & 4.620(3) & 6.731(16) & 7.453(8)\\
  \hline
\end{tabular}
\caption{Critical filling factors $\nu_c$ extracted from the conductance data of Fig.~\ref{fig:fig1}.}
\label{tab:critical_filling}
\end{table}

\subsection{Validity of the analysis}

We have systematically analyzed the temperature dependence of $\Gbulk$ for all filling factors using several different frameworks. The low-temperature data was analyzed through the rescaled conductance plots, illustrated in Fig.~\ref{fig:Hohlsscaling}, based on both ES and Mott-VRH (see below). The non-rescaled conductance was fitted using only VRH (both ES and Mott) at low temperature ($T<2~$K), and a combination of VRH (either ES or Mott) and Arrhenius activation for the full temperature range. The systematic comparison between the different fits is shown in Supplemental Section~II.D~\cite{SM}, showing that the data (both low temperature and full range) is generally better fitted with ES-VRH. Note that spin gaps ($\nu=\pm4,\pm8$) showed a peculiar trend where the low-temperature data was equally well fitted by ES and Mott-VRH (see also below for the comparison between rescaled conductance collapse with ES and Mott-VRH for spin gaps). Importantly, the $T_0$ extracted from the various fitting procedures all follow the same trend (albeit with different orders of magnitude), with the largest $T_0$ obtained for cyclotron gaps, and the smallest for valley gaps~\cite{SM}, in agreement with the observations and interpretations presented here.

\subsection{Rescaled conductance for the $4\rightarrow3$ and $8\rightarrow7$ transitions: Mott versus ES VRH.}

\begin{figure}[ht!]
    \centering
    \includegraphics[width=0.9\linewidth]{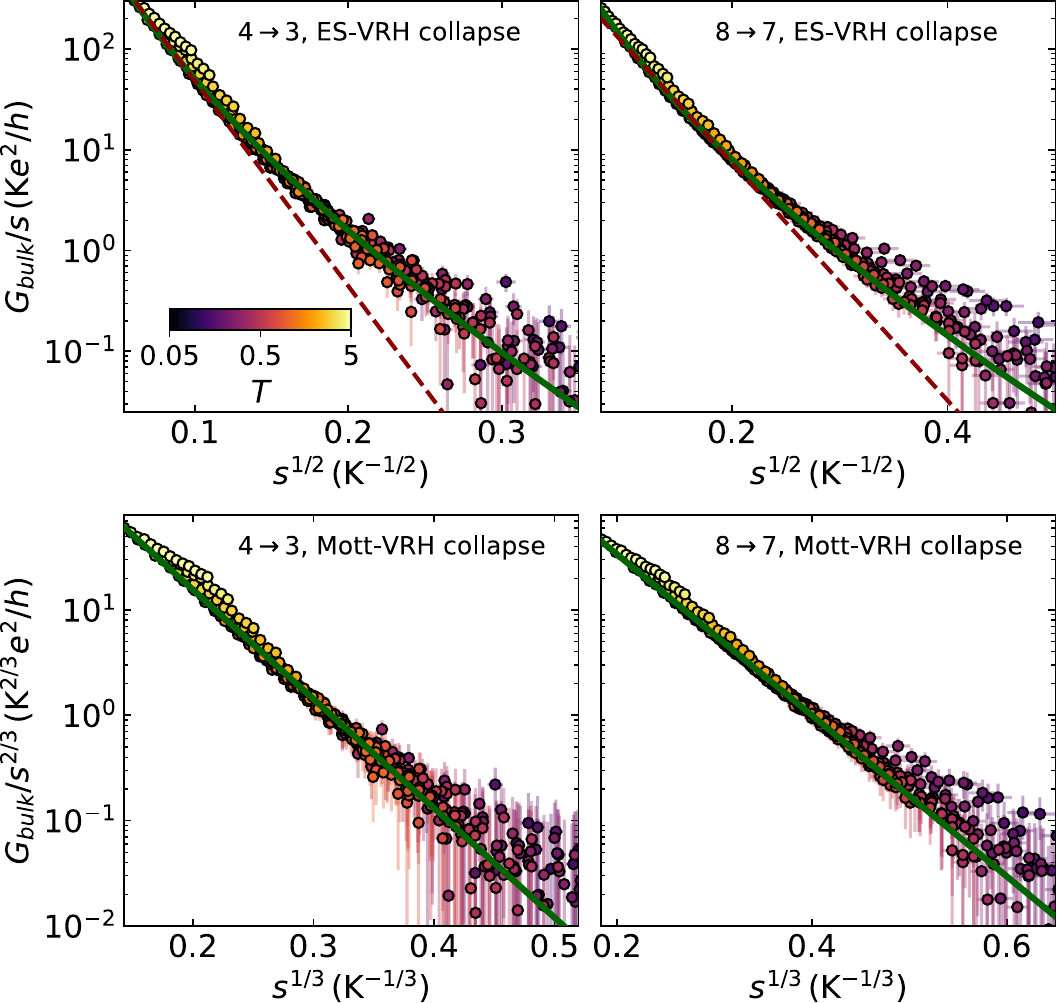}
    \caption{Rescaled conductance data at the $\nu=4\rightarrow3$ (left panels) and $8\leftrightarrow7$ (right panels) transitions for $T<5~$K. The green curves are guides to the eye following Mott's law of VRH (hence straight lines in the lower panels), while the red dashed lines are guides following Efros-Shklovskii's law of VRH.}
    \label{fig:transition_MottVRH}
\end{figure}

In our analysis, most of the conductance data at fixed density as a function of temperature are best fit with the ES-VRH, which accurately describes VRH in presence of a Coulomb gap caused by strong electronic interactions \cite{Shklovskii1984}. Likewise, if the $G_\mathrm{bulk}/s$ data versus $s^{1/2}$ collapse, they are expected to do so on a straight line if they satisfy ES-VRH. However, the data may indeed collapse if the standard criticality is satisfied, while being more accurately described by a non-interacting model of VRH, the so-called Mott VRH. The conductance within that description writes:
\begin{equation}
G_\mathrm{bulk}\propto\left(\frac{T_0}{T}\right)^{2/3}e^{-(T_0/T)^{1/3},}
    \label{EMeq:Mott_VRH}
\end{equation}
where the 1/3 exponent is valid for non-interacting 2D electrons (see the calculation in Supplemental Material Section IV for the temperature dependence of the prefactor), and where the $T_0$ dependence in the prefactor follows the collapse hypothesis.
Interestingly, we identified two regions, across the density/filling factor sweep, where Mott-VRH description was better suited, namely the $4\rightarrow3$ and $8\rightarrow7$ transitions. For these two transitions, we indeed observed convincing data collapse below 1 K, as shown in Fig. \ref{fig:transition_MottVRH} upper panels. However, using the reduced representation $G_\mathrm{bulk}/s$ versus $s^{1/2}$ we found out that the data do not collapse on the straight line that signals ES-VRH (red dashed lines in Fig. \ref{fig:transition_MottVRH} upper panels), with a significant deviation well outside experimental uncertainties. 

Instead, the collapse is well captured by the following expression, where $y=G_\mathrm{bulk}/s$ and $x=s^{1/2}$:
\begin{equation}
    y=Ax^{-2/3}e^{-Bx^{2/3}},
\label{EMeq:modified_collapse}
\end{equation}
which is nothing else than Mott's VRH law (\ref{EMeq:Mott_VRH}), represented by the green solid line in Fig. \ref{fig:transition_MottVRH} upper panels. When using the alternative representation $G_\mathrm{bulk}/s^{2/3}$ versus $s^{1/3}$, we indeed find that the data collapse on a straight line as expected for Mott-VRH, as shown in Fig. \ref{fig:transition_MottVRH} lower panels. Concomitantly, we found out that for the $\Gbulk(T)$ data at fixed filling
$\nu$, Mott VRH is indeed more accurate in these two regions (see Supplemental Material \cite{SM}).
\\
Further investigations, outside the scope of this work, are needed in order to understand this feature. We note, however, that the two transitions that agree with the Mott description are the same within the sub-Landau level hierarchy (from the central spin gap to the lowest valley gap), only for different Landau levels $N=1$ and 2.

\subsection{Extracted activation gaps}

\begin{figure}[ht!]
    \centering
    \includegraphics[width=0.9\linewidth]{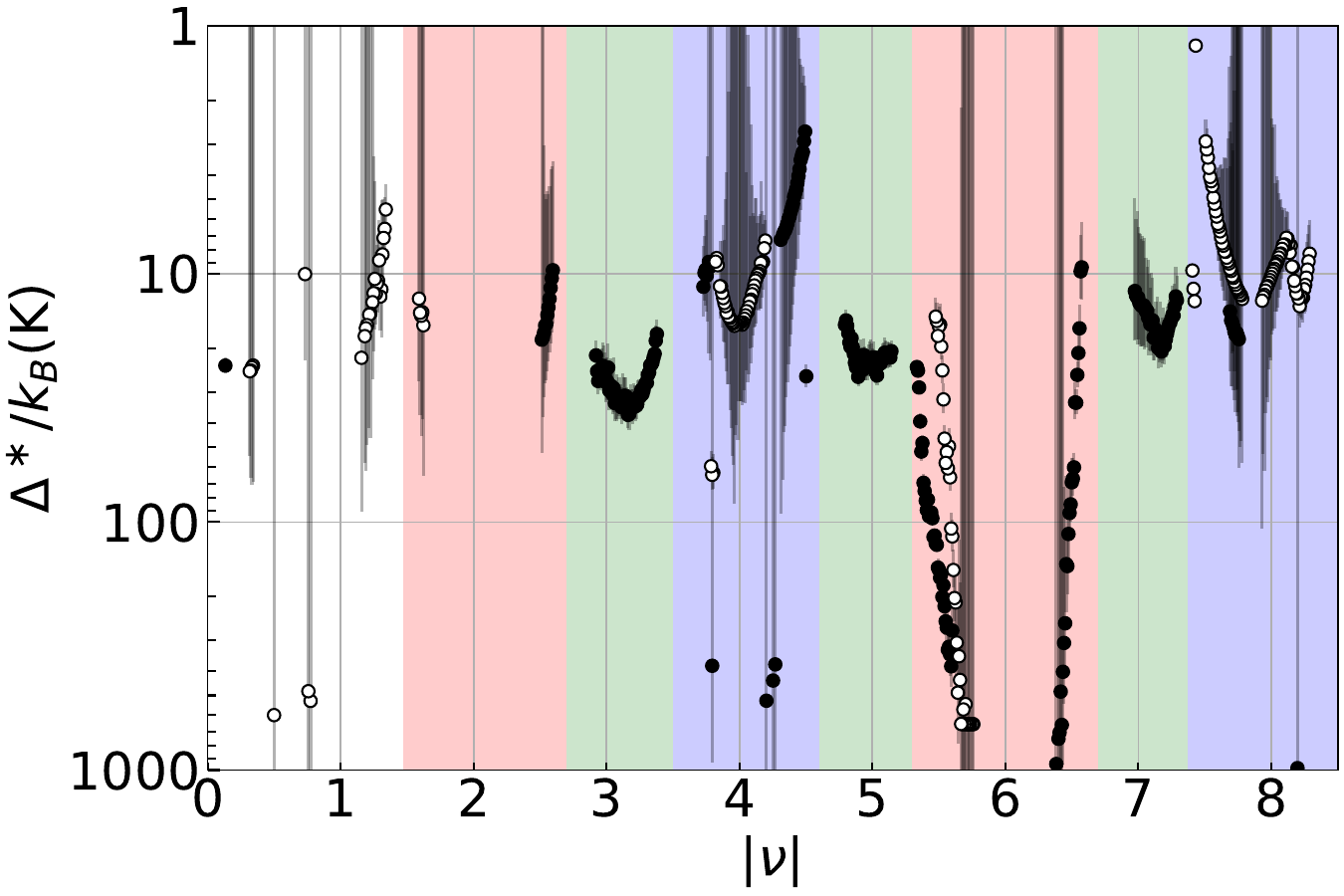}
    \caption{Activation gaps $\Delta^\ast/k_B$ extracted using the ES VRH combined with Arrhenius activation model as a function of the absolute value of filling factors $|\nu|$.}
    \label{fig:activation_gap_14T_ES+activation}
\end{figure}

Using ES VRH combined with Arrhenius activation as fitting model, we have extracted the activation gaps $\Delta^\ast/k_B$ in Fig.~\ref{fig:activation_gap_14T_ES+activation}. For the valley gaps $\nu=3,5,7$, the extracted values correspond to previous observations~\cite{Young2012}, while for cyclotron gaps $\nu=\pm2,\pm6$, we extracted a gap in the order of $100\,\mathrm{K}$ since the experimental temperature is lower than the theoretical gap sizes.

For the spin gaps $\nu=\pm4,\pm8$, the fitting with ES VRH and Arrhenius activation is not as good for other plateaus (see Supp. Mat. \cite{SM} Section II. D), and $\Delta^\ast/k_B\sim10\,\mathrm{K}$. 

\end{document}